Computer-aided Recognition and Assessment of a Porous Bioelastomer on Ultrasound Images for Regenerative Medicine Applications

**Dun Wang[a], Kaixuan Guo[a], Yanying Zhu[a], Jia Sun[a], Aliona Dreglea[b], Jiao Yu[a,1]**

[a]College of Science, Liaoning Petrochemical University, Fushun, Liaoning Province, 113001, P. R. China

[b]Industrial Mathematics Laboratory, Baikal School of BRICS, Irkutsk National Research Technical University, Irkutsk, 664074, Russia

[1] Corresponding author. Address: College of Science, Liaoning Petrochemical University, Fushun, Liaoning Province, 113001, P. R. China. Email: yujiao@lnpu.edu.cn. Tel.: 86-24-56865706. Fax: 86-24-56860766.

Abstract

Biodegradable elastic scaffolds have attracted more and more attention in the field of soft tissue repair and tissue engineering. These scaffolds made of porous bioelastomers support tissue ingrowth along with their own degradation. It is necessary to develop a computer-aided analyzing method based on ultrasound images to identify the degradation performance of the scaffold, not only to obviate the need to do destructive testing, but also to monitor the scaffold's degradation and tissue ingrowth over time. It is difficult using a single traditional image processing algorithm to extract continuous and accurate contour of a target like a porous bioelastomer due to the fuzzy boundary and complex background with a lot of speckle noise in the ultrasound image. To solve these problems, this paper proposes a joint algorithm for the bioelastomer's contour detection and a texture feature extraction method for monitoring the degradation performance of the bioelastomer. The joint algorithm firstly performs manual segmentation on the obtained original ultrasound image to obviate the interference of redundant tissue texture on subsequent detection outcomes. Then mean-shift clustering method is used to obtain the bioelastomer's and native tissue's clustering feature information of the manually segmented image, which is set as the initial value in the image binarization algorithm for image partitioning. Image binarization method adopts the OTSU approach which automatically selects the optimal threshold value to convert the grayscale ultrasound image into a binary image, and black pixels are counted for the area analysis of the bioelastomer. Finally, the Canny edge detector is used to extract the complete bioelastomer's contour from the binary image. Texture feature extraction is based on the computer-aided recognition of the bioelastomer. The region of interest (ROI) is set in the scaffold's region, and the first-order statistical features (mean value, standard deviation, and coefficient of variation) and the second-order statistical features (contrast, entropy, energy, and inverse differential moment) of the grayscale values of ROI are extracted and analyzed. The proposed joint algorithm not only achieves the ideal extraction of the bioelastomer's contours in ultrasound images, but also gives valuable feedback of the degradation behavior of the bioelastomer at the implant site based on the changes of texture characteristics and contour area. The preliminary results of this study suggest that the proposed computer-aided image processing techniques have values and potentials in the non-invasive analysis of tissue scaffolds *in vivo* based on ultrasound images and may help tissue engineers evaluate the tissue scaffold's degradation and cellular ingrowth progress and improve the scaffold designs.

Keywords: ultrasound imaging, edge detection, image processing, texture feature, computer-aided recognition

1 Introduction

In recent years, with the rapid development of computer technology and big data processing, it has become a trend to apply digital image processing technology to the auxiliary research of biomedical ultrasound images. Digital image processing, as a prime image processing and analysis technology[1,2], has extremely high application values in the biomedical field. Digital image processing can enable the quantitative measurements of objective parameters and statistical indices through various algorithms of image segmentation, feature extraction, and image recognition, etc. It helps improve the research efficiency and diagnostic accuracy of specific sites on ultrasound images. Nowadays, prompt progress has been made applying digital image processing technology to diverse biomedical ultrasound images for preclinical and clinical studies. Jain et al.[3] proposed a region difference filter for liver ultrasound image segmentation, which greatly improved the accuracy of image segmentation. Liu et al.[4] implemented image feature extraction to ultrasound images of breast tumors, achieving a higher target tissue recognition rate. Hsu[5] applied image recognition to cardiac ultrasound images to help improve target tracking accuracy. In view of the fact that the recognition resolution of images by computers is much higher than that of human eyes, the computer can more accurately capture and recognize the subtle changes in ultrasound images after processing the high-dimensional data information, providing researchers with more objective ultrasound imaging features.

Biodegradable elastic scaffolds are playing an increasingly important role in tissue engineering and regenerative medicine.[6,7] Scaffolds made of bioelastomers have different biocompatibility and biodegradation behaviors, and they facilitate cell adhesion, proliferation and differentiation. In the study of biodegradable elastomers, polyurethane-based materials have been proved to have highly adjustable biodegradability and have been used as elastic scaffolds in the field of tissue engineering.[8,9] In order to maintain the integrity of scaffolds, it is necessary to develop a non-invasive method to study the degradation performance of scaffolds. It can not only avoid damage of the scaffold, but also monitor the scaffold's degradation and tissue ingrowth over time. With high resolution and considerable imaging depth, ultrasound imaging has been widely used as a non-invasive imaging tool for biological tissues and organs *in vivo*.[10,11]

In this study, a disc-shaped poly(carbonate urethane) urea bioelastomer (10mm diameter, 3mm height) was prepared using salt leaching. It was implanted into the abdomen of a rat and observed for 12 weeks by ultrasound imaging. The experimental setup was identical to that described in Ref. [9]. Then mean-shift clustering method[12], image binarization method[13] and Canny edge detector[14] were used to extract the bioelastomer's contour and analyze its area to study the biodegradation behaviors of the bioelastomer in vivo. By analyzing the first-order and second-order characteristics of statistics of the region of interest (ROI), texture feature extraction of the bioelastomer was performed to study the changes of texture characteristics which provided feedbacks of the bioelastomer's degradation and cellular ingrowth at the implant site.

2 Method

2.1 Ultrasound image acquisition

A high-frequency ultrasound scanner (Vevo 2100, VisualSonics, Canada) was utilized for the ultrasound scans. The abdomen of an anesthetized supine rat was exposed to a 32MHz linear ultrasound probe (MS-550D). The first ultrasound scan at week 0 was performed 3 days after bioelastomer implantation, and subsequent scans were performed every 4 weeks from the first scan date until week 12.[9] The collected RF data were recorded as ultrasound images of 496 x 369 pixels.

2.2 Recognition of bioelastomer

In this study, the original size of each ultrasound image is 496 x 369 pixels, and the upper left corner of the image is the coordinate origin. Assuming that the positive x direction is eastward, and the positive y direction is downward, we first used cropping boxes (x(0:369), y(0:200)) to manually segment the original images of 0-12 weeks. The purpose of manual segmentation is to ensure the information of the 0-12 week bioelastomer is complete, and to eliminate the interference of the remaining biological tissues on the subsequent image segmentation. After that, mean-shift clustering was performed on the manually segmented images, and the clustering feature information was used as input to apply image binarization. The method adopts a grayscale threshold judgment for image segmentation. Finally, the bioelastomer's contour was extracted by Canny edge detector on the processed binary image to complete the recognition of the bioelastomer on ultrasound images.

2.2.1 Mean-shift clustering algorithm

Mean-shift clustering algorithm is a non-parametric clustering algorithm for estimating gradient of the probability density functions. Assuming there are n coordinate points $x_j$ (j = 1,2,...,n) in the d dimensional space $R^d$, the mean-shift vector of $x_j$ at point x is：

$$M_h(x) = \frac{1}{k}\sum_{x_j \in s_h}(x_j - x) \quad (1)$$

where $s_h$ represents a high-dimensional spherical region with a radius of h. In our image processing studies, it represents a circular region of interest with a radius of h. In Eq. 1, k is the number of all coordinate points $x_j$ that fall into the region of interest $s_h$. In the region of interest $s_h$, the set of y points is：

$$S_h = \{y:(y-x_j)^T(y-x_j) \leq h^2\} \quad (2)$$

In Eq. 1, the shift vector of the coordinate point $x_j$ with respect to the centroid x is ($x_j$-x), and the mean value of the shift vector of the centroid $M_h(x)$ is calculated by estimating the kernel density of the region of interest. The mean of the shift vector $M_h(x)$ always points to the direction of the highest probability density gradient, so the centroid x is iteratively shifted towards the direction of the highest density region until the stop condition is met. Each coordinate point can be used as the centroid for iterative computations. Usually, in the region of interest $s_h$, the closer $x_j$ is to the centroid x, the greater the correlation is with the statistical characteristics of the neighboring points. The clustering group with the highest access frequency of all points in the region of interest is regarded as the clustering group of the current point set $s_h(x)$.

In this paper, the pyrMeanShiftFiltering function in OpenCV was used to perform clustering and filtering operations on the manually segmented images. It can erode small areas of grayscale, and cluster the areas with similar grayscale distributions. We set the spatial window radius to 5, the grayscale window radius to 100, and other parameters to default values. This procedure can provide textural features of the elastomer and native tissues for subsequent image binarization.

2.2.2 Image binarization

Image binarization is to classify the pixels on the image according to the thresholding value to achieve the purpose of dividing the target and background. In other words, in a grayscale image with 256 brightness levels, the gray value of all pixels whose gray level is greater than or equal to the threshold is set to 255, otherwise the gray value is set to 0, so that the target and background of the entire image present a clear black and white effect. The functional expression of threshold segmentation is:

$$F(m,n) = \begin{cases} 255, & f(m,n) \geq T \\ 0, & f(m,n) < T \end{cases} \quad (3)$$

In Eq. 3, m and n are pixel coordinates, and T is the set threshold. There are usually two techniques for the selection of threshold: global thresholding technique and local thresholding technique. In global thresholding, a single threshold is used globally, for the whole image. The Otsu's method, named after Nobuyuki Otsu, is a popular global thresholding technique. The Otsu's thresholding method, as an automatic threshold searching algorithm, can derive the optimal threshold on the grayscale image based on the maximization of between-cluster variance. This paper used the Otsu's thresholding method for image processing.

For an image with a size of M x N and a gray level of H, the number of pixels whose gray level is j is $f_j$, then the total number of pixels is:

$$M \times N = \sum_{j=0}^{H-1} f_j \quad (4)$$

In the total number of pixels M x N, the proportion of $f_j$ is

$$p_j = \frac{f_j}{M \times N} \quad (\sum_{j=0}^{H-1} p_j = 1) \quad (5)$$

The proportion of the number of target pixels to the total number of pixels of the image is

$$w_0 = \sum_{j=0}^{T-1} p_j \quad (6)$$

The proportion of the number of background pixels to the total number of pixels of the image is

$$w_1 = \sum_{j=T}^{H-1} p_j \quad (7)$$

The average gray level of the target is

$$\mu_0 = \frac{\sum_{j=0}^{T-1} j \times p_j}{\sum_{j=0}^{T-1} p_j} \quad (8)$$

The average gray level of the background is

$$\mu_1 = \frac{\sum_{j=T}^{H-1} j \times p_j}{\sum_{j=T}^{H-1} p_j} \quad (9)$$

The average gray level of the entire image is

$$\mu = w_0 \times \mu_0 + w_1 \times \mu_1 = \sum_{j=0}^{H-1} j \times p_j \quad (10)$$

The between-cluster variance G(k) between the target and the background is

$$G(k) = w_0(\mu_0 - \mu)^2 + w_1(\mu_1 - \mu)^2 \quad (11)$$

In Eq. 11, the value of k is 0 to 255. The between-cluster variance between the target and the background is calculated by traversing the range of k values. When the between-cluster variance is the largest, the corresponding k value is the optimal threshold T.

The texture features after mean-shift clustering were passed to the threshold function in OpenCV for image segmentation. The initial threshold T was set to 0, the thresholding method was set to THRESH_OTSU, and other parameters was set to default. The algorithm will automatically select the optimal threshold for binarization. In order to make the edge details of the image clearer, it is necessary to use morphological contour detection method for contour edge processing. A rectangular kernel with a size of 3×3 was used to perform open operation on binary images, and the segmented feature images after morphological image processing were displayed after two iterations.

2.2.3 Canny edge detector

As an edge detection operator that uses a multi-stage algorithm to identify edges in images, Canny edge detector mainly uses non-maximum suppression to process grayscale images of gradient magnitude smoothed by Gaussian filter, and then extracts the complete target contours through double-threshold detection.

The high frequency speckle noise in ultrasound images can be mistakenly detected as edge features. In order to improve the detection accuracy of Canny edge detector, it is necessary to first use Gaussian blur as a low-pass filter to remove high-frequency noise. Secondly, since the binarized image will produce gradient changes at the position of black and white transition, this paper used the Sobel operator with convolution kernel size of 10 to perform gradient operation on the images. The generated gradient graph was processed using non-maximum suppression, and the value with large gradient was reserved as the edge information to be selected for further double-threshold detection. Double-threshold detection determines the maximum and minimum thresholds manually. When the gradient value after non-maximum suppression is greater than the set maximum threshold, it is identified as the real edge, and if it is less than the set minimum threshold, it is changed to 0 pixel value and removed. The gradient value between the double thresholds is used as a potential edge for further judgment. If the potential edge is in the same line as the real edge pixels, it will be connected, otherwise it will be removed. Finally, a complete edge was formed to complete the contour extraction.

In this paper, the Canny function in OpenCV was used to extract the contour of the scaffold edge from binarized ultrasound images. The low threshold, high threshold, and Sobel was set to 140, 280, and 10, respectively.

2.3 Image feature extraction

Texture, as an important characteristic parameter of ultrasound images, is due to the difference in the density distribution of different tissue sites, and many randomly distributed scattering particles are generated inside the tissue. The diameter of these scattering particles is much smaller than the ultrasonic wave length. When ultrasound waves pass through the particles, scattering and reflection are generated which yield different echo signal amplitudes, forming an ultrasound image that can reflect the changes in the scaffold's texture.

2.3.1 Region of interest

In this experiment, we selected two groups of regions of interest (ROI) for texture feature extraction. In the first group, we extracted the scaffold's edge contour based on the Canny edge detector and selected the ROI inside the scaffold contour for feature extraction. The size of the ROI is fixed, and it is a rectangular box of size 135×58 pixels. In the second group, the ultrasound image at week 0 was selected, and ROI extraction features within the tissue and scaffold were selected as controls. The size of the ROI is fixed, and it is a rectangular box of size 75×42 pixels. The scaffold ROI of the second group was located at the center of the ROI of the first group at week 0.

2.3.2 Texture feature extraction

Image texture features mainly include first-order and second-order statistical features. The first-order statistical feature is mainly to count the gray value distribution of the image, including the gray-level mean, standard deviation (SD) and coefficient of variation (CV) in the ROI. The second-order statistical feature is mainly to generate gray-level co-occurrence matrix (GLCM)[16,17]. It consists of the grayscale relationship of two pixels in the image space to form a matrix that can reflect the joint distribution probability of their simultaneous occurrence. The elements in GLCM can reflect the distribution characteristics of locations between pixels with identical or similar brightness and thus reflect changes in texture.

Define a GLCM $G(m,n;d,\theta)$, where m and n represent the position coordinates of the two pixels with the gray levels of m, n. d, θ is the distance and angle relationship between two pixels. In order to simplify the calculation, the original image of 256 gray levels is quantized into a 16 gray level image, and a 16×16 GLCM $G(m,n)$, (m = 1,..., 16; n=1,...,16) is generated. Then, the probability matrix $P(m,n)$ is obtained by normalization:

$$P(m,n) = \frac{G(m,n)}{\sum_{m=1}^{16}\sum_{n=1}^{16} G(m,n)} \quad (12)$$

From the probability matrix P(m,n), the statistical features of GLCM can be extracted, including contrast, entropy, energy, and inverse differential moment. The formulas are as follows:

$$f_1 = \sum_{m=1}^{16}\sum_{n=1}^{16}(m-n)^2 P(m,n) \quad (13)$$

$$f_2 = \sum_{m=1}^{16}\sum_{n=1}^{16} P(m,n)\log P(m,n) \quad (14)$$

$$f_3 = \sum_{m=1}^{16}\sum_{n=1}^{16} P(m,n)^2 \quad (15)$$

$$f_4 = \sum_{m=1}^{16}\sum_{n=1}^{16} \frac{P(m,n)}{1+(m-n)^2} \quad (16)$$

The contrast mainly refers to the degree of the spatial intensity change in the image texture. The entropy refers to the randomness of information contained in an image and reflects the complexity of the image texture. The energy is also called the angular second moment, which reflects the uniformity of the grayscale distribution of the image texture. The Inverse differential moment reflects the clarity and regularity of the image texture. In this paper, PyCharm was used to extract the above 7 first-order and second-order statistical features as the overall texture features of the image, aiming at analyzing the pixel value changes inside the scaffold to characterize the biodegradability of poly(carbonate urethane) urea. The operating system used in this paper is Windows 10 Professional Edition 64-bit operating system, the processor is Intel (R) Core (TM) i5-10400 CPU@2.90GHz, 16GB memory, and the programming language is Python 3.6. The compilation environment is PyCharm Community Edition 2020.3.2.

3 Results
3.1 Scaffold identification

As shown in Fig. 1(a), through the joint processing of mean-shift clustering and image binarization algorithm, the scaffold can be extracted from the complex tissue background, and then the ideal scaffold edge contour can be generated by the Canny edge detector. Experimental results show that the joint algorithm in this paper can achieve ideal segmentation and target edge contour extraction for medical ultrasound images.

Through image binarization, the background part of the image was processed into pure white with a pixel value of 255, and the scaffold part was processed into pure black with a pixel value of 0. Then the number of pure black pixel values was counted by PyCharm, and the change of the contour area of the scaffold was obtained. In the process of image binarization, non-binary transition pixels will be generated in the scaffold's edge contour. The transition pixel can represent the perimeter of the scaffold outline, but considering the irregularity of the scaffold contour, the perimeter of the contour as a reference is of little significance. The change of scaffold area is shown in Fig. 1 (b). It could be seen that the area of the scaffold gradually decreases with time except at week 12, which could explain the biodegradation performance of the scaffold in the rat.

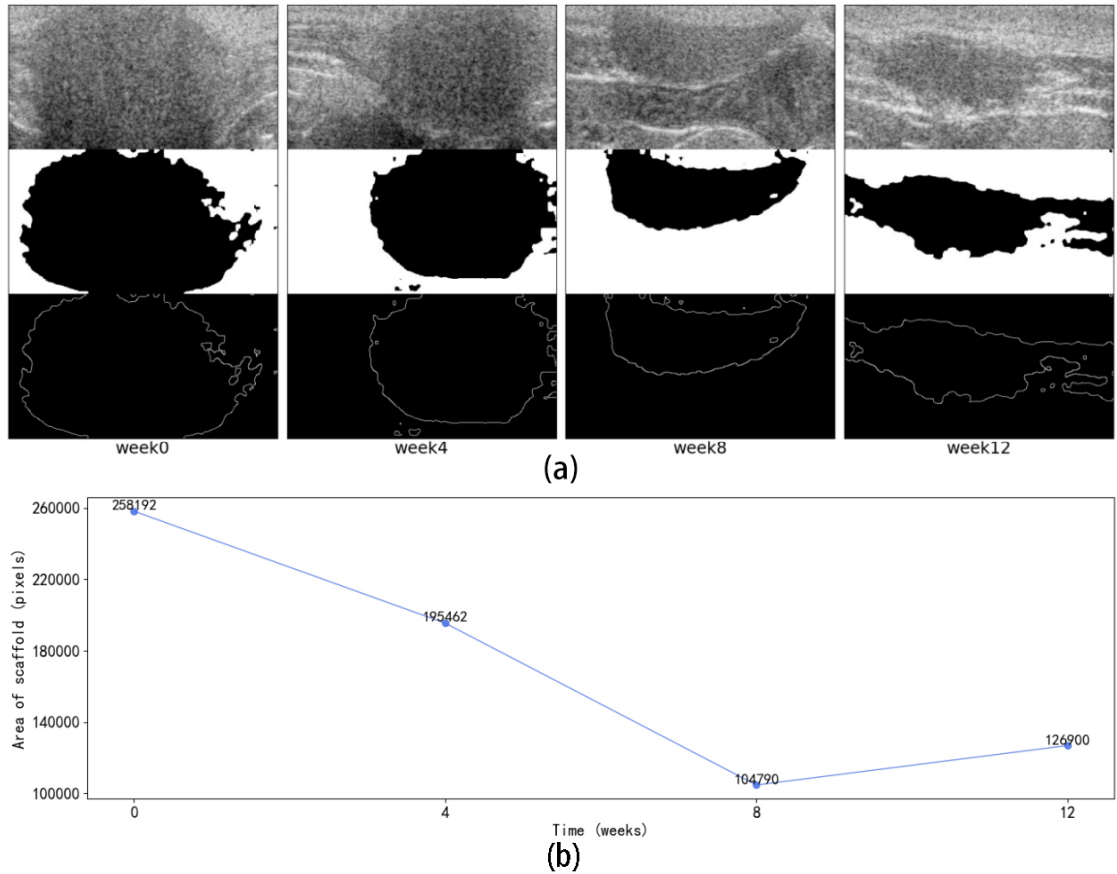

Fig. 1. Scaffold recognition from ultrasound images. (a) The upper part is the ultrasound image after manual segmentation, the middle part is the image after mean-shift clustering and binarization processing, and the lower part is the contour extraction image by using the Canny edge detector. (b) The trend line graph of the change of area of scaffold.

3.2 Results of texture feature analysis

Fig. 2 shows the results of texture feature analysis. In Fig. 2(a), ROIs are marked by blue boxes for the first group and green boxes and red boxes for the second group. From the pixel distribution features extracted from the second group of ROIs in Fig. 2(b), it can be concluded that the overall distribution of pixel values in biological tissues is relatively large. Secondly, through PyCharm, we analyzed that the mean value of pixels in the ROI of tissues was 181.91, which means that a higher pixel value promises a closer characteristics to the surrounding native tissues. Fig. 2(c) shows that the mean pixel value inside the scaffold first decreased and then increased significantly. This indicates that the scaffold first underwent a stage dominated by degradation, during which the ingrowth of cells was not obvious, and then as the tissue ingrowth became dominant, its pixel value continued to increase.

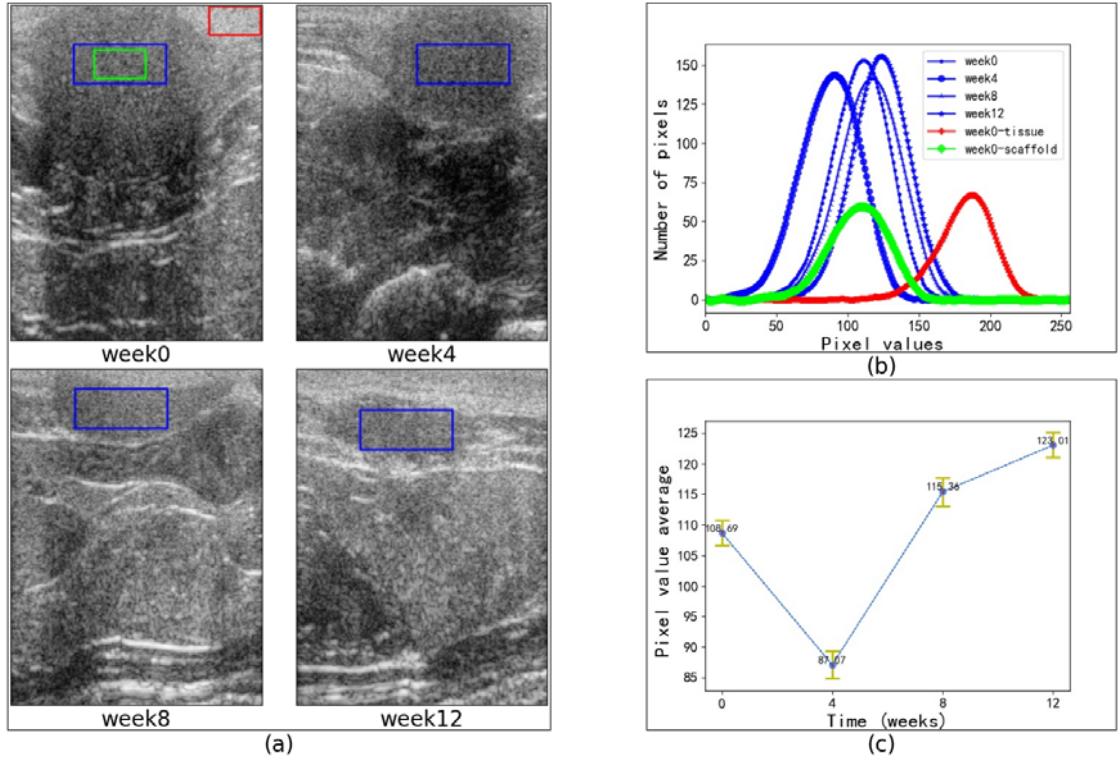

Fig. 2. Results of texture feature analysis. (a) The location of the region of interest. (b) The distribution curve of the pixel values of the scaffold and tissue. (c) The trend line of the average of the pixel value of the scaffold.

The results of the first-order and second-order statistical features are shown in Table 1. According to SD and CV of the first-order statistical features and the contrast and entropy of the second-order statistical features, it can be seen that their changing trends are to increase first, and then to decrease. The changing trend of the inverse differential moment is to first decrease, and then increase. This may be because scaffold degradation and tissue ingrowth are carried out at the same time. Due to the ingrowth of the tissue, there will be both porous scaffold materials with black pixels and tissue contents with white pixels at the scaffold sites. This will change SD and CV of the first-order statistical features and the contrast and entropy of the second-order statistical features. The mixing of black and white pixels will not only increase the overall dispersion of pixel values, but also increase the degree of changes between light and dark in the texture and the complexity of the texture. With the further degradation of the scaffold and further ingrowth of the tissue, the scaffold site is mainly occupied by connective tissues, so the above-mentioned features will tend to gradually decrease. The inverse differential moment is just the opposite, because the inverse differential moment mainly reflects the regularity of the image texture. The ingrowth of the tissue will result in a complex mixed texture pattern of the tissue and the scaffold, and the occupation of the tissue will gradually turn into a single tissue texture pattern. There is no obvious changing trend in energy. In general, the above six statistical features except energy could indicate the degradation and tissue ingrowth of the scaffold in the rat.

Table 1. First-order and second-order statistical features

| TIME | Pixel value average | Standard deviation | Coefficient of variation | Contrast | Entropy | Energy | Inverse differential moment |
|---|---|---|---|---|---|---|---|
| **WEEK 0** | 108.69 | 20.15 | 18.54% | 4.38 | 3.97 | 0.02 | 0.43 |
| **WEEK 4** | 87.07 | 21.16 | 24.30% | 5.64 | 4.23 | 0.02 | 0.40 |
| **WEEK 8** | 115.36 | 21.63 | 18.75% | 4.06 | 3.98 | 0.02 | 0.43 |
| **WEEK 12** | 123.01 | 20.10 | 16.34% | 3.21 | 3.70 | 0.03 | 0.48 |

4 Discussion

In this study, the characterization of the biodegradation performance of a poly(carbonate urethane) urea scaffold was investigated by applying computer-aided recognition method on ultrasound images taken by an ultrasound scanner. Although the ultrasound imaging effect was not ideal due to bubbles when the scaffold was implanted in the rat at week 0, the characteristics of statistics inside the scaffold and the obtained contour area were generally good. The scaffold must have a good biodegradation performance during soft tissue repair to support the adhesion of cells and the ingrowth of tissue. In the ultrasound image of the initial implantation of the scaffold, there are significant differences in the texture of the biological tissues inside and around the scaffold, so the change of the characteristics of statistics of the scaffold can be used to characterize the degradation performance of the scaffold.

In contrast with the changes in the texture features of the scaffold, the changes in the contour area of the scaffold are another manifestation of the biodegradation performance of the scaffold. In this study, mean-shift clustering and image binarization method were used to process the scaffold to pure black. Although there were transition pixels at the edge of the scaffold's contour, the number of black pixels can be used as the area of the scaffold to analyze the biodegradation performance of the scaffold.

Ultrasound imaging has several advantages: it is relatively inexpensive, non-ionizing, portable, efficient and repeatable, and ultrasound imaging enables non-invasive evaluation of tissue, even deep inside the tissue. Although it is still early to know whether ultrasound imaging will prove useful in monitoring scaffold's performance *in vivo*, the results of this preliminary study are very encouraging. To date, efforts devoted to exploring and investigating the potential of ultrasound imaging in characterizing candidate materials for tissue engineering applications appear to be very limited. This paper provides some insights into the *in vitro* characterization of the degradation states of the polyurethane-based elastomer for soft tissue repair by conducting computer-aided analysis of ultrasound images.

The degradation performance of biomaterials is an important index of tissue regeneration based on scaffold substrate, and the appropriate degradation rate should match the growth rate of new tissues, synthetic cells. Bioelastomers that degrade too fast relative to the tissue ingrowth will lose their support functions, leading to premature failure, and bioelastomers that degrade too slow will cause greater mechanical stiffness which tends to yield shear damage to the repaired tissue and cannot achieve the desired repair outcome. Therefore, a balance between the rate of scaffold degradation and tissue formation is very important. Ultrasound imaging combined with computer-aided recognition method is one of the potential directions in meeting the pressing needs for monitoring the degradation rates of the bioelastomers and the increasing amounts of tissue with time, and serves to help tissue engineers develop scaffolds with tailored properties.

The edge detection results have a direct impact on the subsequent image feature extraction and processing. Speckle noise is inherent in ultrasound images, and edge blur and various artifacts make the existing traditional single edge detection algorithms difficult in monitoring accurately the edge of a porous target like a bioelastomer. Therefore, we proposed a joint algorithm for edge detection in this study. On the basis of Canny edge detector, this joint algorithm combines the mean-shift clustering, image binarization and other pre-processing methods, which can achieve a more desirable edge detection of the bioelastomer in ultrasound images. It should be mentioned that we applied the opening operator from the mathematical morphology method after operations of image binarization. The purpose is to effectively smooth the boundary of the bioelastomer without changing its area, so that the extracted scaffold's contour is more clear and continuous. At present, for different medical ultrasound images, there are limitations using one general edge monitoring algorithm, but the joint algorithm proposed in this paper may still be applicable by modifying the built-in parameters, helping carry out contour extraction and area calculation for nonuniform targets, like scaffolds, tumors, etc.

Analysis of the first-order and second-order statistical characteristics of the texture is an important mean for the ultrasonic identification of tissue. Tsuneo et al.[18] extracted the first-order and second-order statistical features of quadriceps femoris in ultrasound images, such as pixel value average, entropy and inverse differential moment, to quantitatively evaluate the morphological changes of aging muscles. Jacinto et al.[19] found that the second-order statistical features could be used as quantitative parameters of muscle ultrasound in clinical diagnosis by extracting the difference between the second-order statistical features of the muscle groups in the upper extremities of patients with muscular atrophy and the control group. In this paper, by extracting and analyzing the changing trend of the first-order and second-order statistical feature information in the two groups of ROIs in ultrasound images, it was found that it can be used as a quantitative index for the qualitative description of the degradation performance of the bioelastomer. Therefore, the texture feature extraction and analysis method proposed in this paper has an important application value in the evaluation of the degradation performance of bioelastomers using ultrasound imaging. It is worth noting that in order to more clearly observe the internal distribution trend of pixel values in the two groups of ROIs, a Savizky-Golay filter was used in this paper to de-noise and smooth the extracted images of pixel value distributions. The biggest characteristics of this filter lies in that the shape and width of the signal can be kept unchanged while the noise is removed.

Although the joint algorithm and the texture feature extraction method proposed in this paper can achieve ideal contour extraction and characterization of the degradation performance of the bioelastomer, there are 3 uncertainties need to be noted. First, the extraction of the scaffold area contour is determined by the threshold value of pixels, which is prone to the interference of surrounding tissues with low gray level and bubbles introduced by the implantation of the scaffold in early stage, and the processing of the edge is not accurate enough. Secondly, the location of ROIs of the elastomer was mainly selected based on the subjective recognition of human eyes so that it was roughly kept in the central position inside the contour of the scaffold, which might introduce some uncertainties in the subsequent analysis results of feature extraction. Finally, in the process of texture feature extraction, we converted the original 256 grayscale ultrasound image into the 16 grayscale image to generate the 16×16 GLCM, and only the grayscale co-occurrence matrix with direction $\theta=0°$ and distance $d=1$ was generated. The second order statistical characteristics obtained by calculating the gray co-occurrence matrix generated by different directions and distances between two pixels are different. Although a group of second-order statistical features can represent the internal variation information of the bioelastomer, the calculations of statistical features of GLCM with multiple directions and distances may be more comprehensive to extract texture features for more objective evaluations.

In this pilot study, we only investigated one sample's characteristics in time using the proposed joint algorithm. Different samples particularly different types of scaffolds may have different biodegradability, and the effectiveness of the experimental method requires more samples to verify. Further research will be conducted to extract and recognize underlying features of scaffold's ultrasound images by using artificial intelligence with deep learning through convolutional neural networks (CNNs), so as to achieve more accurate characterizations of the scaffolds' biodegradation performances and optimize the extraction method of the scaffolds' contours.

5 Conclusion

This study demonstrated that ultrasound imaging combined with computer-aided image processing techniques can be used to evaluate the biodegradability of tissue scaffolds noninvasively in vivo. PyCharm software was used to calculate and analyze the first-order and second-order statistical features of texture inside the scaffold. Mean-shift clustering, image binarization method and Canny edge detector were used to extract the contour and calculate the area of the scaffold. The preliminary results of this study support the potential of computer-aided recognition processing in ultrasound images to non-invasively characterize the biodegradation characteristics of biocompatible polymeric scaffold materials for tissue regeneration or engineering.

ACKNOWLEDGEMENTS
The authors would like to thank Prof. Kang Kim and Prof. Yi Hong who helped in carrying out the research. This work was financially supported by the National Natural Science Foundation of China (12074160), the Natural Science Foundation of Liaoning Province of China (2019-MS-219), and Liaoning Revitalization Talents Program (XLYC1907034).

Reference
[1] Yibiao Rong, Dehui Xiang, Weifang Zhu, Fei Shi, Enting Gao, Zhun Fan, Xinjian Chen. Deriving external forces via convolutional neural networks for biomedical image segmentation. Biomedical Optics Express, 2019, 10(8):3800-3814.

[2] Magda E. Antohe, Doriana A. Forna, Cristina G. Dascalu, Norina C. Forna. Implications of digital image processing in the paraclinical assessment of the partially edentated patient. Revista de Chimie, 2018, 69(2):521-524.

[3] Nishant Jain, Vinod Kumar. Liver ultrasound image segmentation using region-difference filters. Journal of Digital Imaging, 2017, 30(3):376-390.

[4] Yangyang Liu, Li Ren, Xuehong Cao, Ying Tong. Breast tumors recognition based on edge feature extraction using support vector machine. Biomedical Signal Processing and Control, 2020, 58:101825.

[5] Wei-Yen. Hsu. Automatic left ventricle recognition, segmentation and tracking in cardiac ultrasound image sequences. IEEE Access, 2019, 7:140524-140533.

[6] Jarkko P. Hytonen, Jouni Taavitsainen, Santeri Tarvainen, Seppo Yla-Herttuala. Biodegradable coronary scaffolds: their future and clinical and technological challenges. Cardiovascular Research, 2018, 114(8):1063-1072.

[7] Yichuan Pang, An Qin, Xianfeng Lin, Lin Yang, Qiang Wang, Zhengke Wang, Zhi Shan, Shengyun Li, Jiying Wang, Shunwu Fan, Qiaoling Hu. Biodegradable and biocompatible high elastic chitosan scaffold is cell-friendly both in vitro and in vivo. Oncotarget, 2017, 8(22):35583-35591.

[8] Cancan Xu, Yihui Huang, Liping Tang, Yi Hong. Low-initial-modulus biodegradable polyurethane elastomers for soft tissue regeneration. ACS Applied Materials & Interfaces, 2017, 9(3):2169-2180.

[9] Jiao Yu, Keisuke Takanari, Yi Hong, Kee-Won Lee, Nicholas J. Amoroso, Yadong Wang, William R. Wagner, Kang Kim. Non-invasive characterization of polyurethane-based tissue constructs in a rat abdominal repair model using high frequency ultrasound elasticity imaging. Biomaterials, 2013, 34(11):2701-2709.

[10] Rongrong Guo, Guolan Lu, Binjie Qin, Baowei Fei. Ultrasound imaging technologies for breast cancer detection and management: a review. Ultrasound in Medicine and Biology, 2018, 44(1):37-70.

[11] Carmel M. Moran, Adrian J. W. Thomson. Preclinical ultrasound imaging—a review of techniques and imaging applications. Frontiers in Physics, 2020, 8:124.

[12] Jianwu Long, Xin Feng, Xiaofei Zhu, Jianxun Zhang, Guanglei Gou. Efficient superpixel-guided interactive image segmentation based on graph theory. Symmetry-Basel, 2018, 10(5):169.

[13] Zixuan Chen, Xiaodong Lu, Yayu Hao, Zekai Xu, Wenguang Hou, Mingyue Ding. Despeckling of 3D ultrasound medical image on basis of binarization and connectivity. Journal of Medical Imaging and Health Informatics, 2017, 7(3):623-629.

[14] G. Sumathy, J. Arokia Renjit. Distance-based method used to localize the eyeball effectively for cerebral palsy rehabilitation. Journal of Medical Systems, 2019, 43(8):262.

[15] Ta Y. Goh, Shafriza N. Basah, Haniza Yazid, Muhammad J. A. Safar, Fathinul S. A. Saad. Performance analysis of image thresholding: Otsu technique. Measurement, 2018, 114:298-307.

[16] Falguni Chakraborty, Debashis Nandi, Provas K. Roy. Oppositional symbiotic organisms search optimization for multilevel thresholding of color image. Applied Soft Computing Journal, 2019, 82:105577.

[17] Koyuncu Hasan, Mucahid Barstugan. COVID-19 discrimination framework for X-ray images by considering radiomics, selective information, feature ranking, and a novel hybrid classifier. Signal Processing-Image Communication, 2021, 97:116359.

[18] Tsuneo Watanabe, Hiroki MuraKami, Daisuke Fukuoka, Nobuo Terabayashi, Sohee Shin, Tamotsu Yabumoto, Hiroyasu lto, Hiroshi Fujita, Toshio Matsuoka, Mitsuru Seishima. Quantitative sonographic assessment of the quadriceps femoris muscle in healthy Japanese adults. Journal of Ultrasound in Medicine, 2017, 36(7):1383-1395.

[19] Jacinto Javier Martínez-Payá, José Ríos-Díaz, María Elena del Baño-Aledo, Jose Ignacio Tembl-Ferrairó, Juan Francisco Vazquez-Costa, Francesc Medina-Mirapeix. Quantitative muscle ultrasonography using textural analysis in amyotrophic lateral sclerosis. Ultrasonic Imaging, 2017, 39(6):357-368.